\documentstyle[aps,preprint]{revtex}
\input{epsf}
\begin{document}
\draft
\title{ On a Conjecture of Goriely for the Speed of Fronts of the
 Reaction--Diffusion Equation}
\author{J. Cisternas and M.\ C.\ Depassier}
\address{        Facultad de F\'\i sica\\
	P. Universidad Cat\'olica de Chile\\ Casilla 306,
Santiago 22, Chile}

\maketitle
\begin{abstract}
In a recent paper Goriely considers the one--dimensional scalar
reaction--diffusion equation $u_t = u_{xx} + f(u)$ with a polynomial
reaction term $f(u)$ and conjectures the existence of a relation
between a global resonance of the hamiltonian system $ u_{xx} + f(u)
= 0$ and the asymptotic speed of propagation of fronts of the
reaction diffusion equation. Based on this conjecture an explicit
expression for the speed of the front is given. We give a
counterexample to this conjecture and conclude that additional
restrictions should be placed on the reaction terms for which it may
hold.
\end{abstract}

\pacs{82.40.Ck, 47.10+g, 3.40Kf, 2.30.Hq}

\section{Introduction}

The one dimensional scalar reaction-diffusion equation 
\begin{equation}
u_t = u_{xx} + f(u) \label{eq:pde} 
\end{equation}
with $f(0) = f(u_+) = 0$ has been the subject of much study not only
because it models different phenomena
\cite{Britton,Showalter,Volpert}, but also because it is the simplest
reaction-diffusion equation for which rigorous results  can be
obtained \cite{Volpert,KPP37,HR75,Fife1,AW78,Fife2,BD1,BD2,BD3}.
Depending on the situation being considered the reaction term $f(u)$
satisfies additional properties. It has been shown for different
classes of reaction terms that suitable initial conditions $u(x,0)$
evolve in time into a monotonic front joining the state $u=u_+$ to
$u=0$. The asymptotic speed at which the front travels is the minimal
speed for which a traveling monotonic front $u(z) = u( x - c t)$
exists \cite{Fife1,AW78}. Traveling fronts are a solution of the
ordinary differential equation $u_{zz} + c u_{z} + f(u) = 0$. In the
present case we shall be concerned with two types of reaction terms,
the classical case $f > 0 $ in $(0,u_+)$ with $f'(0) > 0$ and the
bistable case $f < 0$ in $(0,a)$, $f>0$ in $(a,u_+)$ with
$\int_0^{u_+} f > 0$ and $f'(0) < 0$. In the classical case there is
a continuum of speeds $c \ge c^*$ for which monotonic fronts exist.
The system evolves into the front of speed $c^*$. In the bistable
case there is a unique isolated value of the speed $c^*$ for which a
monotonic front exists, the system evolves into this front. The
problem is to determine the speed of propagation of the front.  In
the classical case, if in addition $f'(0) > f(u)/u$ the speed of
propagation $c^*$ is the so called linear or KPP value $c_{KPP} = 2
\sqrt{f'(0)}$ \cite{KPP37}. In the other cases (as well as in the
bistable case) there exist variational principles both local and
integral from which the speed can be calculated with any desired
accuracy for arbitrary $f$ \cite{Volpert,HR75,BD2,BD3}.

In a recent paper \cite{Gor95} Goriely proposes a new method for the
determination of the speed. Based on an observed property  of some
exactly solvable cases, namely reaction terms of the form $f(u) = \mu
u + \nu u^n - u^{2 n -1}$ he conjectures that for polynomial reaction
terms of the form
\begin{equation}
f(u) = \mu u + g(u) \qquad 
\label{eq:fhyp}
\end{equation}
 where $g(0) = g'(0) = 0$ and the polynomial $g$ independent of
$\mu$, the speed of the front can be calculated from the knowledge of
the heteroclinic orbit of the Hamiltonian system 
$$ u_{zz} + f(u) = 0.  
$$
This property of solvable cases had not been observed before.
The purpose of this article is to show by means of a counterexample
that this conjecture is not true in general for the above class of
polynomial reaction terms. However, considering that the conjecture
is indeed true for a large  class of reaction terms (the solvable
cases mentioned above), it is a very interesting unsolved problem to
characterize the class of functions for which it is valid.  For the
sake of clarity we state the conjecture here. The conjecture makes
use of the fact the front approaches the equilibrium state $u = 0$ as
${\rm e}^{\lambda_- z}$, a well established fact,  and approaches the
equilibrium point $u= u_+$ as $u = u_+ -   L {\rm e}^{\gamma_+ z}$,
an assumption which is not always satisfied.  Then the global
resonance, defined as
\begin{equation}
\delta = -{\gamma_+ \over \lambda_-} \label{eq:delta}
\end{equation}
is conjectured to be a constant, for a general class of polynomial
reaction terms, at all values of $\mu$ for which the nonlinear front
exists.  Explicit expressions are known for the rates of approach
$\gamma_+$ and $\lambda_-$ in terms of $c$ and $f$ therefore, if
$\delta$ can be calculated at any point, then an analytic formula for
the speed can be obtained.  There is such a point where it can be
calculated, and that is the point at which $c=0$ and the system is
hamiltonian. There is a unique value $\mu < 0$ for which such a front
exists, (we shall label it as $\mu_h$ and the corresponding
equilibrium point as $u_h$)  and therefore the speed is completely
determined.  In the following section we consider a specific
polynomial reaction term and show that it fails to satisfy the
conjecture.

\section{The Counterexample}

Consider the reaction term
$
f(u) = \mu u + 2 u^2 - 7 u^3 + {20\over 3} u^4 - 2 u^5.
$
This is of the form $f(u) = \mu u + g(u)$ where the  polynomial $g(u)
=  2 u^2 - 7 u^3 + 20 u^4/3 - 2 u^5$ satisfies the properties $g(0) =
g'(0) = 0$ and is independent of $\mu$ as requested by the
conjecture. For the hamiltonian system 
$$
 u_{zz} +   \mu u + 2 u^2 - 7 u^3 + {20\over 3} u^4 - 2 u^5 = 0 
$$
 a heteroclinic orbit joining
two equilibrium points exists at the value $\mu = \mu_h = -0.153897$.
In Fig. 1 the reaction term $f(u)$ is shown together with the
(scaled) potential. It is clear that a heteroclinic solution joining
the point $u = 0$ to $ u_+ = u_h = 0.262156$ exists. In this case the
resonance  $\delta$ can be calculated. Its value is given by
\begin{equation}
\delta = \delta_h = \sqrt{{f'(u_h)\over \mu_h} } = 0.865558. \label{eq:deltah}
\end{equation}
Let us now consider the propagating fronts which are a solution of 
$$
u_{zz} +  c u_z +  \mu u + 2 u^2 - 7 u^3 + {20\over 3} u^4 - 2 u^5 = 0.
$$
Before giving the results of the numerical and analytical
calculations we show the plot of the function $f$ at several values
of $\mu$ which will make clear the numerical and analytical results
that follow. As $\mu$ increases the equilibrium point $u_+$ increases
until at $\mu = 1/3$ it reaches the value $u_+ = 1$ where $f' = 0$.
Above this value of $\mu$ there is a discontinuous jump in $u_+$, the
front joins the origin $u = 0$ to a new fixed point which corresponds
to a different root of the polynomial $f(u)$. In Fig. 2 we show the
function $f$  at different values of $\mu$. At $\mu = 1/3$ the fixed
point $u_+ = 1$ and the derivative $f'(u_+) = 0$. At $\mu = 0.4$, we
see that the value of $u_+$ is now the new root of $f$ which did not
exist at low values of $\mu$.

First we describe the results of the numerical integrations of the
initial value problem for Eq(\ref{eq:pde}) with sufficiently
localized initial value perturbations $u(x,0)$. The speed is obtained
numerically and the value of $\delta$ is then computed from
Eq.(\ref{eq:delta}) which can be expressed as
\cite{Gor95}
\begin{equation}
\delta = {-c + \sqrt{c^2 - 4 f'(u_+)} \over c + \sqrt{c^2 - 4 f'(0)} }. 
\label{eq:fordel}
\end{equation}
In Fig. 3 we show the asymptotic speed of the front as a function of
$\mu$. The solid line gives the numerical results and the dashed line
corresponds to the linear or KPP value $2 \sqrt{\mu}$. First we
observe that the KPP value is lower than the calculated speed in the
range of $\mu$ shown which means that the transition to the linear or
KPP regime occurs at larger values of $\mu$. Even though the value of
$u_+$ is discontinuous, the speed is a continuous function of $\mu$.
 
 And finally the graph of the resonance $\delta$ as a function of
$\mu$ is shown in Fig. 4. At $\mu = \mu_h$ it adopts the analytically
calculated value from the hamiltonian case, decreases to a value
$\delta = 0$ at $\mu = 1/3$ jumps discontinuously to a larger value
and increases from there on.  This discontinuity can be attributed to
the discontinuity in $u_+$. At the value of $\mu = 1/3$ where
$f'(u_+) = 0$ it is evident from Eq (\ref{eq:fordel}) that $\delta =
0$.  As we will show below, at this value of $\mu$ the speed and the
asymptotic behavior for the front can be calculated analytically and
it is found that the front does not approach the fixed point $u_+ =
1$ exponentially, therefore one of the assumptions of the conjecture
does not hold. Indeed we shall show that at $\mu = 1/3$ the front
approaches $u= 1$ as $ u \sim 1 - A/z$.

In order to determine the speed we shall make use of variational
principles.  It is known that the speed of the front is given by
\cite{BD2}
\begin{equation}
c = \max 2 { \int_0^{u_+} \sqrt{ - f \phi \phi'} \,du \over
\int_0^{u_+} \phi \, du} \label{eq:var1}
\end{equation}
where the maximum is taken over positive decreasing functions $\phi
(u)$.  Taking as a trial function 
$$
\phi(u) = { ( 1 -u)^{7/2} \over \sqrt{u} } {\rm e}^{- 3/[(2 (1 -u)]}
$$
 the integrals in Eq.(\ref{eq:var1}) can be performed. We obtain
for $\mu = 1/3$ 
$$
 c \ge \sqrt{{3\over 2}} > 2 \sqrt{\mu}.  
$$
 To
obtain an upper bound we make use of the local variational principle
\cite{HR75} 
$$ c = \inf_\rho \sup_u \left( \rho'(u) + {f(u)\over
\rho(u) }\right) 
$$
 where the trial function $\rho(u) > 0$ and
$\rho'(0) > 0$. Choosing as a trial function $
\rho(u) = \sqrt{ {2/ 3} }\, u \,( 1 - u)^2 
$
we obtain the upper bound 
$$
c\le \sqrt {{3\over 2}}
$$
which combined with the above lower bound implies that the speed is
exactly $c = \sqrt{3/2}$ in agreement with the results of the
numerical integration.  The exact value of the speed could be
obtained analytically from the variational principles due to the fact
that for $\mu = 1/3$ the derivative of the front can be calculated
exactly. The derivative of the front as a function of $u$, $ p (u) =
- d u/d z$ satisfies the equation $ p(u) p'(u) - c p(u) + f(u) = 0$
and the exact solution at $\mu = 1/3$ is given by $ p (u) = \sqrt{
{2/ 3} }\, u\, ( 1 - u)^2$. With this expression for $p$ we may
calculate the approach to the fixed point $u = 1$.  Near $u = 1$, $p
\sim
\sqrt{2/3}\, (1 - u)^2$ so that 
$$
{du \over dz} \sim -\sqrt{{2\over 3}} \,(1 - u)^2
$$
from where it follows that 
$$
 u(z)  \sim 1 - \sqrt{{3\over 2}}\, {1\over z}.
$$ 
We see then that at this point $\delta = 0$ since the rate of
approach is not exponential but algebraic and one of the assumptions
of the conjecture is not satisfied. We conclude from this example
that the conjecture does not hold for general polynomials of the form
given by Eq.(\ref{eq:fhyp}).

\section{Conclusion}

We have seen by means of a counterexample that the conjecture put
forward that relates certain properties of the hamiltonian system
$u_{zz} + f(u) = 0$  with the speed of the front solution of $u_{zz}
+ c u_z + f(u) = 0$ is not satisfied by general polynomial reaction
terms $f(u)$. As observed by Goriely, there is a class of reaction
terms for which it does hold, those for which an exact solution for
the front $u(z)$ can be given explicitly.  Numerical evidence has
been given \cite{Gor95} in at least one case where the conjecture
seems to hold in a case where the front cannot be calculated
explicitly. On the other hand, we have given a counterexample to this
conjecture. It is an interesting problem to establish precisely the
conditions under which the proposed conjecture holds. This would lead
to a classification of systems in at least three classes, those for
which the speed is given in terms of the derivative at one fixed
point, that is the  KPP  value $c = 2\sqrt{f'(0)}$, those in which
the speed would be determined by the derivatives at the two fixed
points (the expression given by Goriely would hold) and the rest, for
which the speed depends on integral properties of the reaction term.

\section{Acknowledgments}
This work was partially supported by Fondecyt project 1960450.

\newpage
\begin{centerline}{
{\bf Figure Captions}}
\end{centerline}

\vspace{1.0cm}
{\bf Figure 1}

Graph of the reaction term $f$ and the corresponding scaled potential at the 
value of $\mu$ for which the speed of the front vanishes and the system is hamiltonian.

\vspace{1.0cm}

{\bf Figure 2}

Graph of the reaction term at different values of $\mu$. The value of the stable point increases with $\mu$ until $\mu$ reaches $1/3$. A discontinuous jump in the stable point occurs at that value.

\vspace{1.0cm}

{\bf Figure 3}

Graph of the speed obtained from the numerical integration of the initial value problem. The speed of the front is a continuous function of $\mu$. In the range of $\mu$ shown the speed is greater than the linear or KPP value.

\vspace{1.0cm}

{\bf Figure 4}

Value of the resonance $\delta$ as a function of $\mu$ obtained from the numerical integrations. 
\newpage
\epsfbox{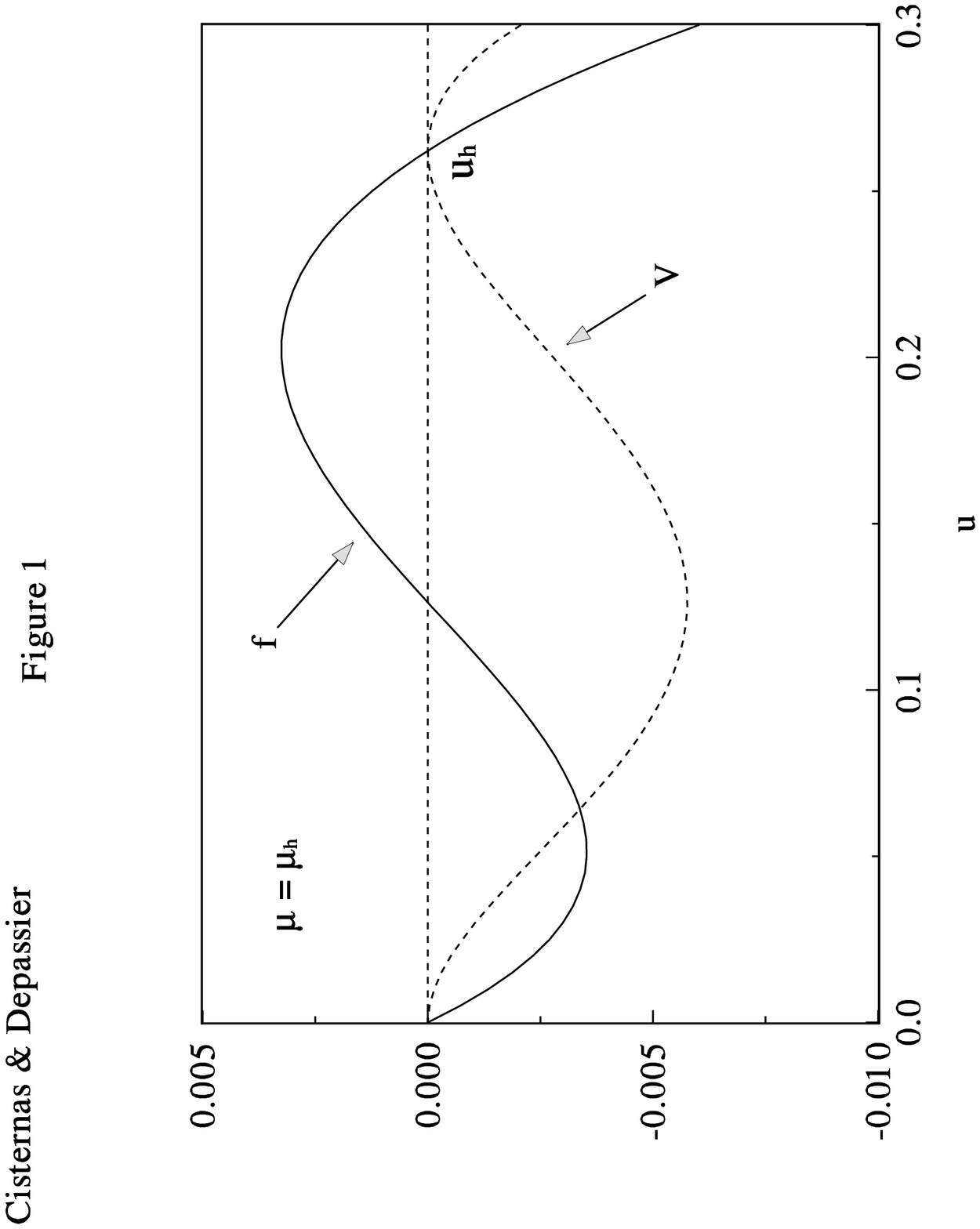}
\newpage
\epsfbox{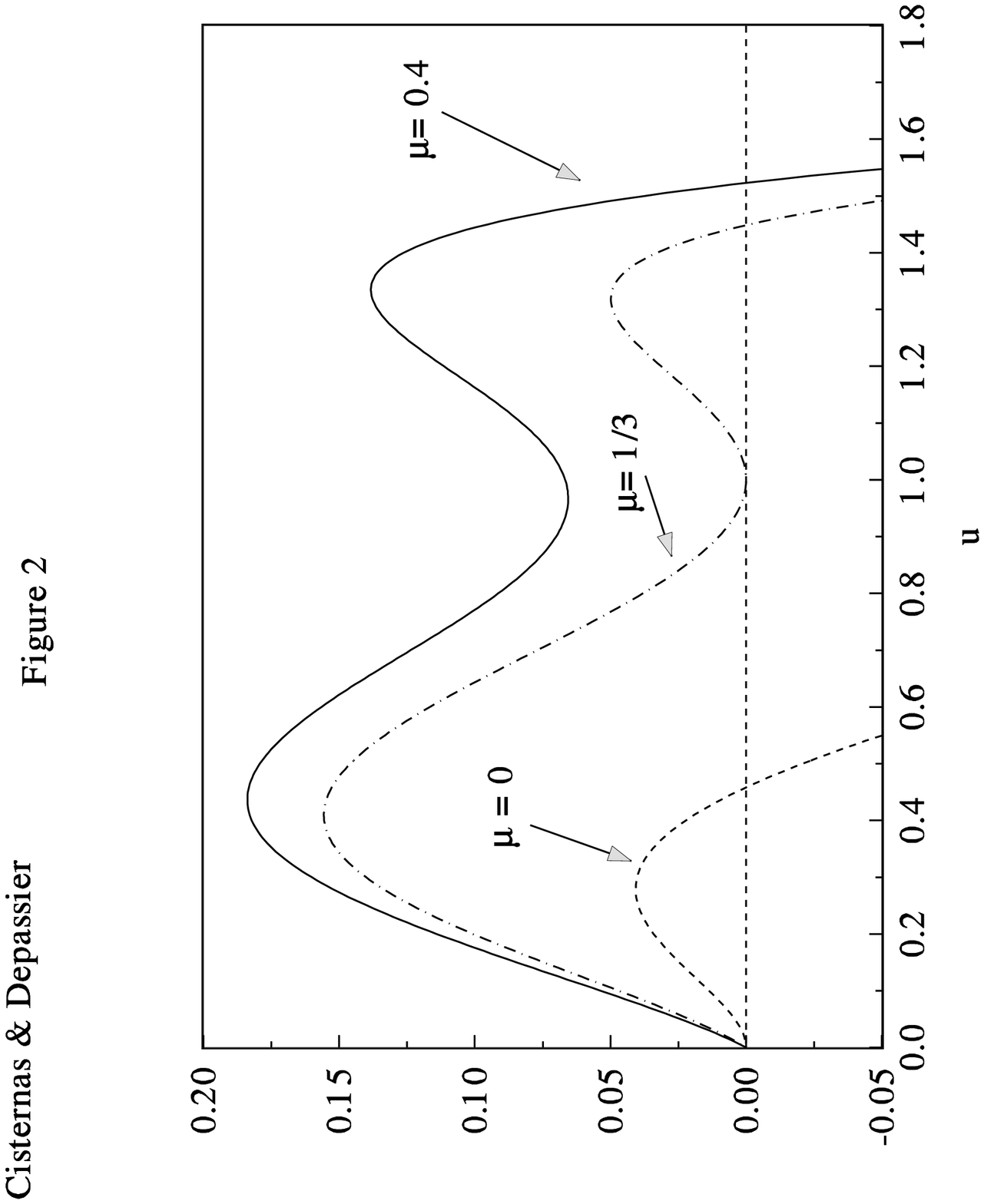}
\newpage
\epsfbox{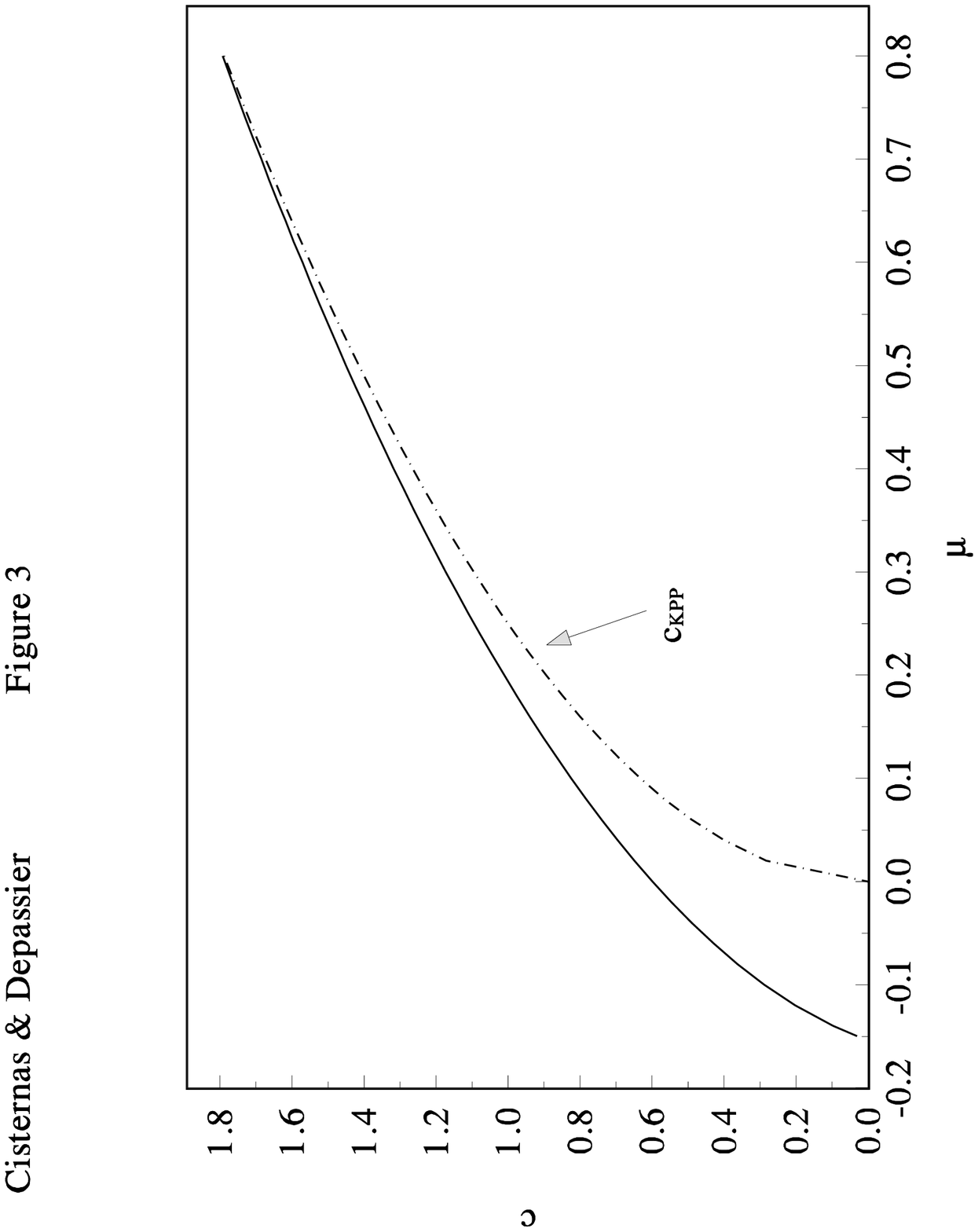}
\newpage
\epsfbox{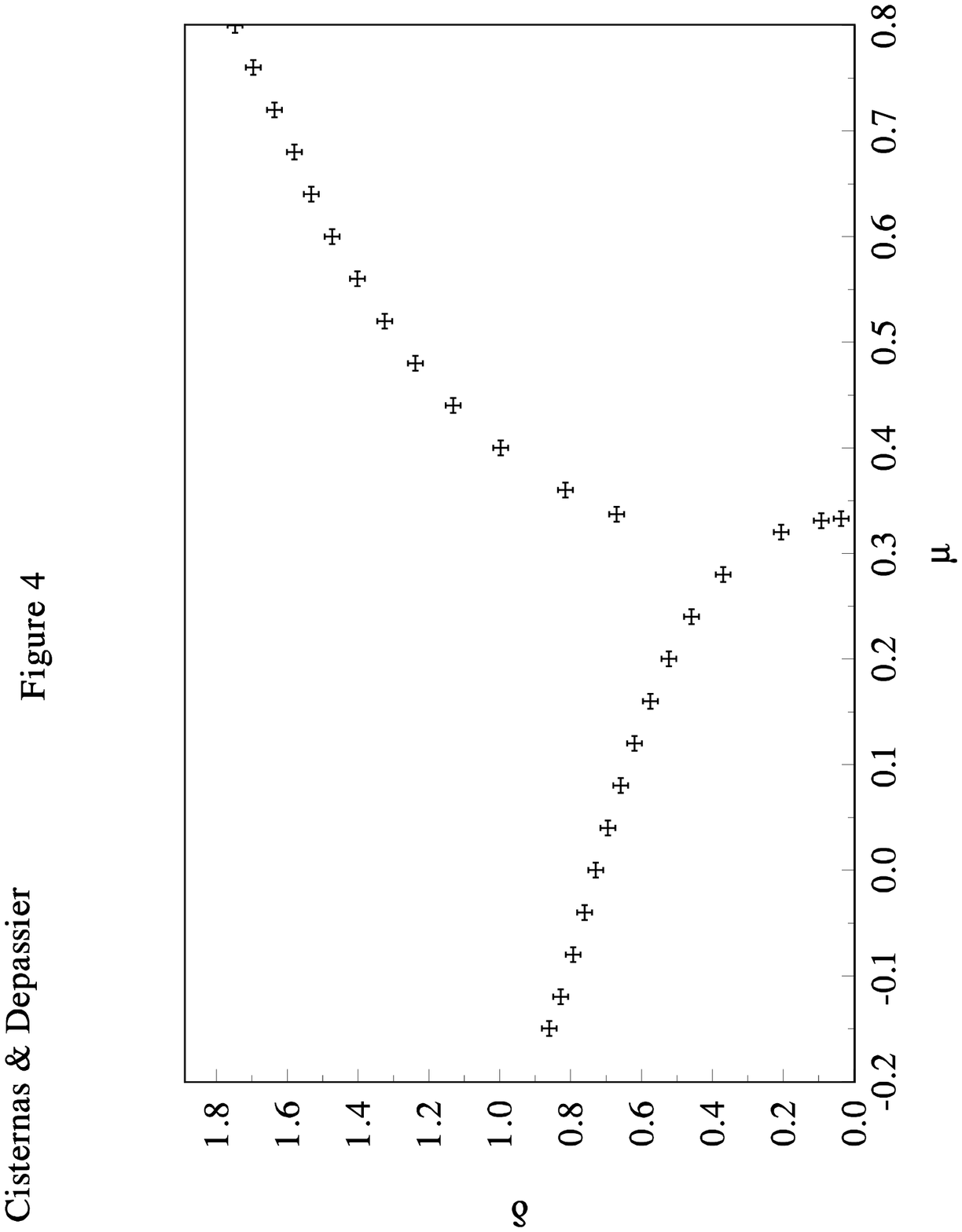}

\end{document}